\documentclass[12pt]{article}
\usepackage{amssymb,amscd,array}
\catcode `\@=11
\@addtoreset{equation}{section}

\newtheorem{thm}{Theorem}[section]
\newtheorem{rem}[thm]{Remark}
\newtheorem{lemma}[thm]{Lemma}

\def\qed{\blacksquare}
\newcommand{\be}{\begin{equation}}
\newcommand{\ee}{\end{equation}}
\newcommand{\bea}{\begin{eqnarray}}
\newcommand{\eea}{\end{eqnarray}}

\newcommand{\N}{\mathbb{N}}

\def\d{\partial}
\begin{document}
\begin{titlepage}

\begin{center}
{\bf \Large{Off-Shell Fields and Quantum Anomalies\\}}
\end{center}
\vskip 1.0truecm
\centerline{D. R. Grigore, 
\footnote{e-mail: grigore@theory.nipne.ro}}
\vskip5mm
\centerline{Department of Theoretical Physics}
\centerline{Institute for Physics and Nuclear Engineering ``Horia Hulubei"}
\centerline{Bucharest-M\u agurele, P. O. Box MG 6, ROM\^ANIA}

\vskip 2cm
\bigskip \nopagebreak
\begin{abstract}
\noindent
We use the formalism of quantum off-shell fields for the case of pure Yang-Mills fields. In this formalism one can compute in a systematic way the second order anomalies of the tree sector. 
\end{abstract}
\end{titlepage}

\section{Introduction}

The general framework of perturbation theory consists in the construction of 
the chronological products such that Bogoliubov axioms are verified \cite{BS}, \cite{EG}, \cite{DF}, \cite{Gl}, \cite{Sc2}, \cite{Sto1}; for every set of Wick monomials 
$ 
W_{1}(x_{1}),\dots,W_{n}(x_{n}) 
$
acting in some Fock space
$
{\cal H}
$
generated by the free fields of the model, one associates the operator
$ 
T_{W_{1},\dots,W_{n}}(x_{1},\dots,x_{n}); 
$  
all these expressions are in fact distribution-valued operators called chronological products. Sometimes it is convenient to use another notation: 
$ 
T(W_{1}(x_{1}),\dots,W_{n}(x_{n})). 
$ 
The construction of the chronological products can be done recursively according to Epstein-Glaser prescription \cite{EG}, \cite{Gl} (which reduces the induction procedure to a distribution splitting of some distributions with causal support) or according to Stora prescription \cite{PS} (which reduces the renormalization procedure to the process of extension of distributions). These products are not uniquely defined but there are some natural limitation on the arbitrariness. An equivalent point of view uses retarded products \cite{St1}.

Gauge theories describe particles of higher spin. Usually such theories are not renormalizable. However, one can save renormalizability using ghost fields. Such theories are defined in a Fock space
$
{\cal H}
$
with indefinite metric, generated by physical and un-physical fields (called {\it ghost fields}). One selects the physical states assuming the existence of an operator $Q$ called {\it gauge charge} which verifies
$
Q^{2} = 0
$
and such that the {\it physical Hilbert space} is by definition
$
{\cal H}_{\rm phys} \equiv Ker(Q)/Ran(Q).
$
The space
$
{\cal H}
$
is endowed with a grading (usually called {\it ghost number}) and by construction the gauge charge is raising the ghost number of a state. Moreover, the space of Wick monomials in
$
{\cal H}
$
is also endowed with a grading which follows by assigning a ghost number to every one of the free fields generating
$
{\cal H}.
$
The graded commutator
$
d_{Q}
$
of the gauge charge with any operator $A$ of fixed ghost number
\be
d_{Q}A = [Q,A]
\ee
is raising the ghost number by a unit. It means that
$
d_{Q}
$
is a co-chain operator in the space of Wick polynomials. From now on
$
[\cdot,\cdot]
$
denotes the graded commutator.
 
A gauge theory assumes also that there exists a Wick polynomial of null ghost number
$
T(x)
$
called {\it the interaction Lagrangian} such that
\be
~[Q, T(x)] = i \partial_{\mu}T^{\mu}(x)
\label{gauge-1}
\ee
for some other Wick polynomials
$
T^{\mu}.
$
This relation means that the expression $T$ leaves invariant the physical states, at least in the adiabatic limit. Indeed, one can write the preceding identity as
\be
~[Q, T(f)] = - i~T^{\mu}(\partial_{\mu}f)
\label{gauge-1a}
\ee
where $f$ is a test function. So, when this test function becomes flatter and flatter we have 
\be
~[Q, T(f)] \approx 0
\ee
so the interaction Lagrangian leaves invariant the physical states.

In all known models there exists a chain of Wick polynomials
$
T^{\mu},~T^{\mu\nu},~T^{\mu\nu\rho},\dots
$
such that:
\be
~[Q, T] = i \partial_{\mu}T^{\mu}, \quad
[Q, T^{\mu}] = i \partial_{\nu}T^{\mu\nu}, \quad
[Q, T^{\mu\nu}] = i \partial_{\rho}T^{\mu\nu\rho},\dots
\label{descent}
\ee
In all cases
$
T^{\mu\nu},~T^{\mu\nu\rho},\dots
$
are completely antisymmetric in all indices; it follows that the chain of relation stops at the step $4$ (if we work in four dimensions). We can also use a compact notation
$
T^{I}
$
where $I$ is a collection of indices
$
I = [\nu_{1},\dots,\nu_{p}]~(p = 0,1,\dots,)
$
and the brackets emphasize the complete antisymmetry in these indices. All these polynomials have the same canonical dimension
\be
\omega(T^{I}) = \omega_{0},~\forall I
\ee
and because the ghost number of
$
T \equiv T^{\emptyset}
$
is supposed null, then we also have:
\be
gh(T^{I}) = |I|.
\ee
One can write compactly the relations (\ref{descent}) as follows:
\be
d_{Q}T^{I} = i~\partial_{\mu}T^{I\mu}.
\label{descent1}
\ee

For concrete models the chain of descent equations (\ref{descent}) can stop earlier. We can construct the chronological products
$$
T^{I_{1},\dots,I_{n}}(x_{1},\dots,x_{n}) \equiv T(T^{I_{1}}(x_{1}),\dots,T^{I_{n}}(x_{n}))
$$
according to the recursive procedure. We say that the theory is gauge invariant in all orders of the perturbation theory if the following set of identities generalizing (\ref{descent1}):
\be
d_{Q}T^{I_{1},\dots,I_{n}} = 
i \sum_{l=1}^{n} (-1)^{s_{l}} {\partial\over \partial x^{\mu}_{l}}
T^{I_{1},\dots,I_{l}\mu,\dots,I_{n}}
\label{gauge}
\ee
are true for all 
$n \in \N$
and all
$
I_{1}, \dots, I_{n}.
$
Here we have defined
\be
s_{l} \equiv \sum_{j=1}^{l-1} |I|_{j}.
\ee
In particular, the case
$
I_{1} = \dots = I_{n} = \emptyset
$
it is sufficient for the gauge invariance of the scattering matrix, at least
in the adiabatic limit; this can be argued as in (\ref{gauge-1a}).

Such identities can be usually broken by {\it anomalies} i.e. expressions of the type
$
A^{I_{1},\dots,I_{n}}
$
which are quasi-local and might appear in the right-hand side of the relation (\ref{gauge}). One compute these anomalies in lower orders of perturbation theory and imposing their cancellation one obtains various restrictions on the expression of the interaction Lagrangian. In this paper we consider the interaction between pure Yang-Mills fields. We compute the anomalies of this model in the second order of perturbation theory using the formalism of off-shell fields. This formalism give a systematic way of computing the anomalies. Such a formalism was used previously in the literature in the context of classical field theory. We use here a pure quantum version.

In the next Section we remind our definition of free fields. We avoid explicit formulas using the reconstruction theorem of Wightmann. In Section \ref{int} we recall the main result concerning the interaction Lagrangians for the most simple model with higher spin fields, namely the pure Yang-Mills fields model. In Section \ref{off} we  introduce the off-shell formalism. In Section \ref{pert} we consider perturbative quantum field theory in the second order. Then in Section \ref{second} we describe the consequences of the cancellation of the anomalies in the second order of the perturbation theory.

\section{Free Fields\label{free}}

We will adopt the description of free quantum fields given by the reconstruction theorem from axiomatic field theory \cite{J}, \cite{SW} based on Borchers algebras. In this approach one can construct a quantum field giving the Wightmann $n$-points distributions and the statistics. For a free field it is sufficient to give the Wightmann $2$-points distribution and generate the rest according to Wick theorem. We use formal distribution notations for simplicity.

\subsection{\bf The Real Scalar Field\label{scalar}} 

We start with the most elementary case of a real scalar field. The field is
$
\Phi(x)
$
and the Hilbert space is generated by vectors of the type
\be
\Phi(x_{1}) \cdots \Phi(x_{n})~\Omega
\label{states-scalar}
\ee
where 
$
\Omega
$
is the vacuum vector. By definition, the $2$-points distribution is
\be
<\Omega, \Phi(x_{1}) \Phi(x_{2})\Omega> = - i~D_{m}^{(+)}(x_{1} - x_{2})
\label{2-scalar}
\ee
where
$
D_{m}^{(+)}(x)
$
is the positive frequency part of the Pauli-Jordan causal distribution of mass $m$. We assume that the scalar field is a Bose field and the $n$-points distributions are generated according to Wick theorem: for $n$ odd
\be
<\Omega, \Phi(x_{1}) \cdots \Phi(x_{n})\Omega> = 0
\label{n-odd-scalar}
\ee
and for $n$ even:
\bea
<\Omega, \Phi(x_{1}) \cdots \Phi(x_{n})\Omega> = 
\sum_{\sigma} <\Omega, \Phi(x_{\sigma(1)}) \Phi(x_{\sigma(2)})\Omega> \cdots
<\Omega, \Phi(x_{\sigma(n-1)}) \Phi(x_{\sigma(n)})\Omega>;
\label{n-even-scalar}
\eea
here the sum is over all permutations $\sigma$ of the numbers
$
1, 2, \dots, n.
$
We also postulate that the field 
$
\Phi
$
is self-adjoint:
\be
\Phi^{\dagger} = \Phi.  
\label{adj-scalar}
\ee

Then one can construct the Hilbert space 
$
{\cal H}
$
from vectors of the type (\ref{states-scalar}) with the scalar product 
$
<\cdot,\cdot>
$
reconstructed from the $n$-points distributions given above and the self-adjointness assumption. We first define a sesquilinear form in the Hilbert space between two states of the form (\ref{states-scalar}) by
\bea
< \Phi(x_{n})\cdots \Phi(x_{1})\Omega, \Phi(x_{n+1})\cdots \Phi(x_{m+n})\Omega>
\nonumber \\
\equiv < \Omega, \Phi(x_{1})^{\dagger} \cdots \Phi(x_{n})^{\dagger} \Phi(x_{n+1})\cdots \Phi(x_{m+n})\Omega>
= < \Omega, \Phi(x_{1}) \cdots \Phi(x_{m+n})\Omega>
\label{scalar-product}
\eea
and one can prove that is positively defined so it induces a scalar product. 

Then the action of the scalar field on states of the form (\ref{states-scalar}) is defined in an obvious way. One can prove that the scalar field so defined verifies the Klein-Gordon equation of mass $m$
\be
K_{m} \Phi = (\square + m^{2}) \Phi = 0; \qquad 
\square \equiv \partial^{2} = \partial\cdot\partial = \partial_{\mu}~\partial^{\mu}
\label{KG-scalar-eq}
\ee
and the canonical commutations relation:
\be
[\Phi(x_{1}), \Phi(x_{2}) ] = - i~D_{m}(x_{1} - x_{2}).
\ee

Because of this commutation relation the writing of a state from the Hilbert space in the form (\ref{states-scalar}) is not unique.

Moreover, one can introduce in the Hilbert space
$
{\cal H}
$
a unitary (irreducible) representation of the Poincar\'e group according to
\bea
U_{\Lambda,a} \Phi(x) U_{\Lambda,a}^{-1} = \Phi(\Lambda^{-1}\cdot (x - a)) 
\nonumber \\
U_{\Lambda,a}~\Omega = \Omega
\label{representation-scalar}
\eea
(here 
$
\Lambda \in {\cal L}^{\uparrow}_{+}
$ 
is a proper orthochronous Lorentz transform and $a$ is a space-time translation). One can obtain in an elementary way the action of the operator 
$
U_{\Lambda,a}
$
on vectors of the type (\ref{states-scalar}) by commuting the operator with the factors
$
\Phi(x_{j})
$
till it hits the vacuum and gives the identity. In the same way one can define the space and time parity operators.

Of course, one can obtain very explicit representations for the scalar field (see e.g. \cite{We}, but they will be not needed in the following. We only mention that one can define in the same way the Wick (or normal) products
$
:\Phi^{n}(x):
$
for any integer $n$ (see \cite{WG}).

For an ensemble of real scalar fields
$
\Phi_{a},~a = 1,\dots,r
$
we only replace (\ref{2-scalar}) by
\be
<\Omega, \Phi_{a}(x_{1}) \Phi_{b}(x_{2})\Omega> = 
- i~\delta_{ab}~D_{m}^{(+)}(x_{1} - x_{2})
\label{2-scalars}
\ee 
and we make a corresponding modification of the formula (\ref{n-even-scalar}). A complex scalar will be an appropriate combination of two real scalar fields. For Fermi fields, the signature of the permutation should be introduced in formulas of the type (\ref{n-even-scalar}).

\subsection{Yang-Mills Fields\label{yang-mills}}

For fields of higher spin one can use the preceding formalism with one major modification: it is necessary to introduce ghosts fields, which are fields with the ``wrong statistics".

The generic case is the Massless vector field. In this case we consider the vector space 
$
{\cal H}
$
of Fock type generated (in the sense of Borchers theorem) by the following fields:
$
(v^{\mu}, u, \tilde{u})
$
where the non-zero $2$-point distributions are
\bea
<\Omega, v^{\mu}(x_{1}) v^{\nu}(x_{2})\Omega> = 
i~\eta^{\mu\nu}~D_{0}^{(+)}(x_{1} - x_{2}),
\nonumber \\
<\Omega, u(x_{1}) \tilde{u}(x_{2})\Omega> = - i~D_{0}^{(+)}(x_{1} - x_{2}),
\qquad
<\Omega, \tilde{u}(x_{1}) u(x_{2})\Omega> = i~D_{0}^{(+)}(x_{1} - x_{2}).
\label{2-massless-vector}
\eea

We also assume the following self-adjointness properties:
\be
v_{\mu}^{\dagger} = v_{\mu}, \qquad 
u^{\dagger} = u, \qquad
\tilde{u}^{\dagger} = - \tilde{u}.
\label{adj-vector-null}
\ee

When we generate the $n$-point functions according to a formula of the type (\ref{n-odd-scalar}) and (\ref{n-even-scalar}) we assume that the field 
$
v^{\mu}
$
is Bose and the fields
$
u, \tilde{u}
$
are Fermi. When defining the unitary representation of the Lorentz group we consider that
the first field is vector and the last two are scalars. Because of the ``wrong" statistics the sesquilinear form defined by a formula of the type (\ref{scalar-product}) will not be positively defined. Nevertheless, because it is non-degenerated, we can prove that we have Klein-Gordon equations of null mass:
\be
\square~v^{\mu} = 0 \qquad \square u = 0 \qquad \square \tilde{u} = 0
\label{KG-vector-null-eq}
\ee
and the canonical commutations relation:
\bea
[v^{\mu}(x_{1}), v^{\nu}(x_{2}) ] = i~\eta^{\mu\nu}~D_{0}(x_{1} - x_{2})
\nonumber \\
\{ u(x_{1}), \tilde{u}(x_{2}) \} = - i~D_{0}(x_{1} - x_{2})
\label{CCR-vector-null}
\eea
and all other (anti)commutators are null.

We can obtain a {\it bona fid\ae} scalar product introducing the so-called {\it gauge charge} i.e. an operator $Q$ defined by:
\bea
~[Q, v^{\mu}] = i~\partial^{\mu}u,\qquad
\{ Q, u \} = 0,\qquad
\{Q, \tilde{u}\} = - i~\partial_{\mu}v^{\mu}
\nonumber \\
Q \Omega = 0.
\label{Q-vector-null}
\eea
Using these relation one can compute the action of $Q$ on any state generated by a polynomial in the fields applied on the vacuum by commuting the operator $Q$ till it hits the vacuum and gives zero. However, because of the canonical commutation relations the writing of a polynomial state is not unique. One can prove that the operator $Q$ leaves invariant the canonical (anti)commutation relations given above and this leads to the consistency of the definition. Then one shows that the operator $Q$ squares to zero:
\be
Q^{2} = 0
\ee 
and that the factor space
$
Ker(Q)/Ran(Q)
$
is isomorphic to the Fock space particles of zero mass and helicity $1$ (photons and gluons) \cite{cohomology}.

We can generalize this case considering the tensor product of $r$ copies of massless vector fields, i.e. we consider the set of fields
$
(v^{\mu}_{a}, u_{a}, \tilde{u}_{a}),~a = 1,\dots,r
$
of null mass and we extend in an obvious way the definitions of the scalar product and of the gauge charge.

\section{Interactions\label{int}}

The discussion from the Introduction provides the physical justification for determining the cohomology of the operator 
$
d_{Q} = [Q,\cdot]
$
induced by $Q$ in the space of Wick polynomials. A polynomial 
$
p \in {\cal P}
$
verifying the relation
\be
d_{Q}p = i~\d_{\mu}p^{\mu}
\label{rel-co}
\ee
for some polynomials
$
p^{\mu}
$
is called a {\it relative co-cycle} for 
$
d_{Q}.
$
The expressions of the type
\be
p = d_{Q}b + i~\d_{\mu}b^{\mu}, \qquad (b, b^{\mu} \in {\cal P})
\ee
are relative co-cycles and are called {\it relative co-boundaries}. We denote by
$
Z_{Q}^{\rm rel}, B_{Q}^{\rm rel} 
$
and
$
H_{Q}^{\rm rel}
$
the corresponding cohomological spaces. In (\ref{rel-co}) the expressions
$
p_{\mu}
$
are not unique. It is possible to choose them Lorentz covariant. We have a general description of the most general form of the interaction of the previous fields \cite{cohomology}. Summation over the dummy indices is used everywhere. For simplicity we do not write the double dots of the Wick product notations.

\begin{thm}
Let $T$ be a relative co-cycle in the variables 
$
(v^{\mu}_{a}, u_{a}, \tilde{u}_{a}),~a = 1,\dots,r
$
which is tri-linear in the fields, of canonical dimension
$
\omega(T) \leq 4
$
and ghost number
$
gh(T) = 0.
$
Then:
(i) $T$ is (relatively) cohomologous to a non-trivial co-cycle of the form:
\be
t = f_{abc} \left( {1\over 2}~v_{a\mu}~v_{b\nu}~F_{c}^{\nu\mu}
+ u_{a}~v_{b}^{\mu}~\d_{\mu}\tilde{u}_{c}\right)
\label{int-ym}
\ee

(ii) The relation 
$
d_{Q}t = i~\d_{\mu}t^{\mu}
$
is verified by:
\be
t^{\mu} = f_{abc} \left( u_{a}~v_{b\nu}~F^{\nu\mu}_{c} -
{1\over 2} u_{a}~u_{b}~\d^{\mu}\tilde{u}_{c} \right)
\label{t-mu-ym}
\ee

(iii) The relation 
$
d_{Q}t^{\mu} = i~\d_{\nu}t^{\mu\nu}
$
is verified by
\be
t^{\mu\nu} \equiv {1\over 2} f_{abc}~u_{a}~u_{b}~F_{c}^{\mu\nu}.
\label{t-munu-ym}
\ee
and we have
$
d_{Q}t^{\mu\nu} = 0.
$

(iv) The constants
$
f_{abc}
$
must be completely antisymmetric
\be
f_{abc} = f_{[abc]}
\label{anti-f}
\ee
and the expressions given above are self-adjoint iff the constants
$
f_{abc}
$
are real. Here we have defined the gauge invariants which are not coboundaries
\be
F^{\mu\nu}_{a} \equiv \d^{\mu}v^{\nu}_{a} - \d^{\nu}v^{\mu}_{a}, 
\quad \forall a = 1,\dots,r
\ee 
\label{t-ym}
\end{thm}

There are different ways to obtain the preceding results. One can proceed by brute force, making an ansatz for the expressions
$
T^{I}
$
and solving the identities of the type (\ref{descent1}) as it is done in \cite{Sc2}. There are some tricks to simplify such a computation. The first one makes an ansatz for 
$
T
$
and eliminates the most general relative cocycle. Then one computes 
$
d_{Q}T
$
and writes it as a total divergence plus terms without derivatives on the ghost fields. Another trick is to use the so-called descent procedure. The first line of proof starts from the general form:
\bea
T = f^{(1)}_{abc} v_{a}^{\mu} v_{b}^{\nu} \partial_{\mu}v_{c\mu} 
+ f^{(2)}_{abc} v_{a}^{\mu} v_{b\mu} \partial_{\nu}v_{c}^{\nu} 
+ f^{(3)}_{abc}~\epsilon_{\mu\nu\rho\sigma}~v_{a}^{\mu} v_{b}^{\nu} \partial^{\sigma}v_{c}^{\rho}
\nonumber \\
+ g^{(1)}_{abc} v^{\mu}_{a} u_{b} \partial_{\mu}\tilde{u}_{c}
+ g^{(2)}_{abc} \partial_{\mu}v^{\mu}_{a} u_{b} \tilde{u}_{c}
+ g^{(3)}_{abc} v^{\mu}_{a} \partial_{\mu}u_{b} \tilde{u}_{c}.
\eea
Eliminating relative coboundaries we can fix: 
\be
f^{(1)}_{abc} = - f^{(1)}_{bac},\qquad
f^{(2)}_{abc} = 0,\qquad
g^{(3)}_{abc} = 0,\qquad
g^{(2)}_{abc} = g^{(2)}_{bac}.
\ee
Then we obtain easily:
\be
d_{Q}T = i u_{a} T_{a}  + {\rm total~div}
\ee
where:
\bea
T_{a} = - 2 f^{(1)}_{abc}~\partial^{\nu}v^{\mu}_{b}~\partial_{\mu}v_{c\nu}
+ (f^{(1)}_{cba} + g^{(2)}_{bac})~
\partial_{\mu}v^{\mu}_{b}~\partial_{\nu}v_{c}^{\nu}
\nonumber \\
+ (- f^{(1)}_{abc} + f^{(1)}_{cba} + f^{(1)}_{bca} + g^{(1)}_{bca})~
v^{\mu}_{b}~\partial_{\mu}\partial_{\nu}v^{\nu}_{c}
\nonumber \\
- 2 f^{(3)}_{abc}~\epsilon_{\mu\nu\rho\sigma}
~\partial^{\mu}v^{\nu}_{b}~\partial_{\sigma}v_{c\rho}.
\eea

Now the gauge invariance condition (\ref{gauge-1}) becomes
\be
u_{a} T_{a} = \partial_{\mu}t^{\mu}
\label{gauge-11}
\ee
for some expression
$
t^{\mu}
$
which has, from power counting arguments, the general form
\be
t^{\mu} = u_{a}~t^{\mu}_{a} + \partial^{\mu}u_{a}~t_{a}
+ \partial_{\nu}u_{a}~t^{\mu\nu}_{a}
\ee
where the polynomial
$
t^{\mu\nu}_{a}
$
does not contain terms with the factor 
$
\eta^{\mu\nu}.
$
Then the relation (\ref{gauge-11}) is equivalent to:
\bea
\d_{\mu}t^{\mu}_{a} - m_{a}^{2}~t_{a} = T_{a}
\nonumber \\
t^{\mu}_{a} + \d^{\mu}t_{a} + \d_{\nu}t^{\nu\mu}_{a} = 0
\nonumber \\
t^{\mu\nu}_{a} = t^{\nu\mu}_{a}.
\eea
One can obtain easily from this system that
\be
T_{a} = (\square + m_{a}^{2})~t_{a}.
\ee
Writing a generic form for 
$
t_{a}
$
it is easy to prove that in fact:
\be
T_{a} = 0;
\ee
from here we easily obtain the total antisymmetry of the expressions
$
f^{(1)}_{abc}
$
and
$
f^{(3)}_{abc};
$
also we have
$
g^{(2)}_{abc} = 0.
$
Now one can take 
$
f^{(3)}_{abc} = 0
$
if we subtract from $T$ a total divergence. As a result we obtain the (unique) solution:
\be
T = f^{(1)}_{abc} ( v_{a}^{\mu} v_{b}^{\nu} \partial_{\nu}v_{c\mu}
- v_{a}^{\mu} u_{b} \partial_{\mu}\tilde{u}_{c})
\ee
which is the expression from the theorem.

Now we briefly present the descent method in this case. There are two results which must be used repeatedly \cite{cohomology}. First, we have a version of the Poincar\'e lemma valid for Wick monomials and then we have a description of the cohomology group
$
H_{Q}
$
of
$
d_{Q}
$
in terms of invariants: if $T$ is a Wick polynomial verifying
$
d_{Q}~T = 0
$
then it is of the form 
$
T = d_{Q}B + T_{0}
$
where 
$
T_{0}
$
depends only on the gauge invariants
$
u_{a}, F_{a}^{\mu\nu}.
$

By hypothesis we have
\be
d_{Q}T = i~\d_{\mu}T^{\mu}.
\label{descent-a}
\ee
If we apply 
$
d_{Q}
$
we obtain
$
\d_{\mu}d_{Q}~T^{\mu} = 0.
$
Using Poincar\'e lemma one finds out some Wick polynomials
$
T^{[\mu\nu]}
$
such that
\be
d_{Q}T^{\mu} = i~\d_{\nu}T^{[\mu\nu]}.
\label{descent-b}
\ee
Continuing in the same way we find
$
T^{[\mu\nu\rho]}
$
such that
\be
d_{Q}T^{[\mu\nu]} = i~\d_{\rho}T^{[\mu\nu\rho]};
\label{descent-c}
\ee
we also have
\be
gh(T^{I}) = |I|.
\ee 

It means that
$
T^{[\mu\nu\rho]}
$
is a sum of terms of the type
$
\eta^{\mu\nu}~u_{a}~u_{b}~\d^{\rho}u_{c}
$
i.e. is a coboundarco-boundary
$
T^{[\mu\nu\rho]} = d_{Q}B^{[\mu\nu\rho]}.
$
We introduce in (\ref{descent-c}) and obtain 
\be
d_{Q}(T^{[\mu\nu]} - i~\d_{\rho}B^{[\mu\nu\rho]}) = 0.
\ee
Using the description of the cohomology of
$
d_{Q}
$
we can easily find that we have:
\be
T^{[\mu\nu]} = d_{Q}B^{[\mu\nu]}+ i~\d_{\rho}B^{[\mu\nu\rho]} + T^{[\mu\nu]}_{0}
\ee 
where the last term depends only on the invariants 
$
u_{a}, F_{a}^{\mu\nu}
$
i.e.
\be
T_{0}^{[\mu\nu]} = {1\over 2}~f^{(1)}_{[ab]c}~u_{a}~u_{b}~F_{c}^{\mu\nu}
+ {1\over 2}~f^{(2)}_{[ab]c}~\epsilon^{\mu\nu\rho\sigma}~u_{a}~u_{b}~F_{c\rho\sigma};
\ee
We substitute these expressions in (\ref{descent-b}) and obtain 
\be
d_{Q}(T^{\mu} - i~\d_{\nu}B^{[\mu\nu]} - t^{\mu}) = 0
\ee
where:
\be
t^{\mu} \equiv f^{(1)}_{[ab]c}~\left(u_{a}~v_{b\nu}~F_{c}^{\nu\mu}
- {1\over 2}~u_{a}~u_{b}~\d^{\mu}\tilde{u}_{c}\right)
- f^{(2)}_{[ab]c}~\epsilon^{\mu\nu\rho\sigma}~u_{a}~v_{b\nu}~F_{c\rho\sigma}.
\ee

If we use again the cohomology of
$
d_{Q}
$
we can easily find out that in fact:
\be
T^{\mu} = d_{Q}B^{\mu} + i~\d_{\nu}B^{[\mu\nu]} + t^{\mu}.
\ee
We substitute this in (\ref{descent-a}) and we obtain the restrictions
\be
f^{(1)}_{[ab]c} = - f^{(1)}_{[ac]b}, \qquad f^{(2)}_{[ab]c} = - f^{(2)}_{[ac]b}
\nonumber
\ee
so the constants 
$
f^{(1)}_{[ab]c},~f^{(2)}_{[ab]c}
$
are in fact completely antisymmetric and
\be
d_{Q}(T^{\mu} - i~\d_{\nu}B^{\mu\nu} - t) = 0
\ee
where
\be
t \equiv f^{(1)}_{[abc]} \left( {1\over 2}~v_{a\mu}~v_{b\nu}~F_{c}^{\nu\mu}
+ u_{a}~v_{b}^{\mu}~\d_{\mu}\tilde{u}_{c}\right)
- {1\over  2}~f^{(2)}_{[abc]}~
\epsilon_{\mu\nu\rho\sigma}~v_{a}^{\mu}~v^{\nu}_{b}~F^{\rho\sigma}_{c}. 
\ee
The description of the cohomology of
$
d_{Q}
$
leads to
\be
T = d_{Q}B + i~\d_{\mu}B^{\mu} + t.
\ee
Finally one proves that the last term from the expression $t$ is a total divergence.

\section{Perturbation Theory\label{pert}}
Now we proceed to the (second order) of perturbation theory. Our purpose is to compute the scattering matrix
\be
S(g) \equiv I + i \int dx g(x) T(x) 
\nonumber \\
+ {i^{2}\over 2} \int dx~dy~g(x)~g(y)~T(x,y) + \cdots 
\label{S-matrix}
\ee
where $g$ is some test function. The expressions 
$T(x,y)$
are called {\it (second order) chronological products} because they must verify the causality property:
\be
T(x,y) = T(x) T(y)
\ee
for 
$x \succ y$
i.e. 
$(x - y)^{2} \geq 0, x^{0} - y^{0} \geq 0$;
in other words the point $x$ succeeds causally the point $y$. This is some generalization of the property
\be
U(t,s) = U(t,r) U(r,s),~~t > r > s
\ee
of the time evolution operator from non-relativistic quantum mechanics.

We go to the second order of perturbation theory using the {\it causal commutator}
\be
D^{A,B}(x,y) \equiv D(A(x),B(y)) = [ A(x),B(y)]
\ee
where 
$
A(x), B(y)
$
are arbitrary Wick monomials. These type of distributions are translation invariant i.e. they depend only on 
$
x - y
$
and the support is inside the light cones:
\be
supp(D) \subset V^{+} \cup V^{-}.
\ee

A theorem from distribution theory guarantees that one can causally split this distribution:
\be
D(A(x),B(y)) = A(A(x),B(y)) - R(A(x),B(y)).
\ee
where:
\be
supp(A) \subset V^{+} \qquad supp(R) \subset V^{-}.
\ee
The expressions 
$
A(A(x),B(y)), R(A(x),B(y))
$
are called {\it advanced} resp. {\it retarded} products. They are not uniquely defined: one can modify them with {\it quasi-local terms} i.e. terms proportional with
$
\delta(x - y)
$
and derivatives of it. 

There are some limitations on these redefinitions coming from Lorentz invariance and {\it power counting}: this means that we should not make the various distributions appearing in the advanced and retarded products too singular.

Then we define the {\it chronological product} by:
\be
T(A(x),B(y)) = A(A(x),B(y)) + B(y) A(x) = R(A(x),B(y)) + A(x) B(y).
\ee

The expression
$
T(x, y)
$
corresponds to the choice
\be
T(x,y) \equiv T(T(x), T(x)).
\ee

The ``naive'' definition
\be
T(A(x),B(y)) = \theta(x^{0}-y^{0}) A(x) B(y) + \theta(y^{0}-x^{0}) B(y) A(x)
\ee
involves an illegal operation, namely the multiplication of distributions. This appears in the loop contributions (the famous ultraviolet divergences). 

\newpage 
\section{The Off-Shell Formalism\label{off}}

It is known that in the second order of the perturbation theory some anomalies can appear and this is due essentially because the Pauli-Jordan distribution 
$
D_{m}
$
verifies Klein-Gordon equation:
\be
K_{m}~D_{m} = (\square + m^{2})~D_{m} = 0
\label{kg-d}
\ee
but the associated Feynman distribution
$
D_{m}^{F}
$
verifies
\be
K_{m}~D_{m}^{F} = (\square + m^{2})~D^{F}_{m} = \delta(x - y).
\label{kg-df}
\ee

Let us describe in detail this point. One computes the second order causal commutator and finds out that the tree contribution has the following generic form:
\bea
[T^{I_{1}}(x),T^{I_{2}}(y)]^{\rm tree} = 
\sum_{m}~[~D_{m}(x - y)~A^{I_{1},I_{2}}_{m}(x,y) 
+ \d_{\alpha}~D_{m}(x - y)~A^{I_{1},I_{2};\alpha}_{m}(x,y) ]
\label{T-tree}
\eea
where the sum runs over the various masses from the spectum of the model and the expressions
$
A^{I_{1},I_{2}}_{m}
$
and
$
A^{I_{1},I_{2};\alpha}_{m}
$
are Wick polynomials. Moreover we have from (\ref{descent1}) the identity
\be
d_{Q}[T^{I_{1}}(x),T^{I_{2}}(y)] = i {\d \over \d x^{\mu}}[T^{I_{1}\mu}(x),T^{I_{2}}(y)]
+ (-1)^{|I_{1}|} {\d \over \d y^{\mu}}[T^{I_{1}}(x),T^{I_{2}\mu}(y)]
\label{d-2}
\ee
which stays true if we take only the tree graphs. Now one can find out the corresponding chronological products by simply substituting in the preceding expression the causal distribution by the associated Feynman propagator:
$
D_{m} \rightarrow D_{m}^{F}
$
i.e.
\bea
T^{I_{1},I_{2}}(x,y)^{\rm tree} = 
\sum_{m}~[~D_{m}^{F}(x - y)~A^{I_{1},I_{2}}_{m}(x,y) 
+ \d_{\alpha}~D_{m}^{F}(x - y)~A^{I_{1},I_{2};\alpha}_{m}(x,y) ].
\eea
In this way all Bogoliubov axioms are true (in the second order) but we might break gauge invariance i.e. the identity (\ref{gauge}) for 
$
n = 2
$
\be
d_{Q}T^{I_{1},I_{2}}(x,y) = i {\d \over \d x^{\mu}}T^{I_{1}\mu,I_{2}}(x,y)
+ (-1)^{|I_{1}|} {\d \over \d y^{\mu}}T^{I_{1},I_{2}\mu}(x,y)
\label{t-2}
\ee
might not be true. Indeed, one can find in the chronological product 
$
T^{I_{1}\mu,I_{2}}(x,y)^{\rm tree}
$
terms of the type
$
\d^{\mu}D_{m}^{F}(x - y)~A^{I_{1},I_{2}}(x,y).
$ 
Then, because of the difference between the relations (\ref{kg-d}) and (\ref{kg-df}) we have in the right hand side of (\ref{t-2}) an extra-term 
$
\delta(x - y)~A^{I_{1},I_{2}}(x,y).
$
One must collect all quasi-local terms appearing in this way and check if they can be put under the form of a co-boundary
$
(d_{Q} - i \delta)R^{I_{1},I_{2}}(x,y)
$ 
where 
$
R^{I_{1},I_{2}}(x,y)
$
are quasi-local expressions; then we can restore gauge invariance (at least for the tree contributions) by redefining the chronological products in an obvious way. 

So the first problem is to find out the anomaly i.e. the expression appearing in the right hand side of (\ref{t-2}) and the second problem is to see in which conditions it can be eliminated by a redefinition of the chronological products. Even the first problem is not exactly elementary in complex models as for instance the case of gravity: in \cite{Sc2} one can see for instance that not only terms of the type 
$
\d^{\mu}D_{m}^{F}(x - y)~A(x,y)
$ 
can produce anomalies. So we need a systematic way to compute the anomaly.

This suggests to make the following change in the description of the fields from section \ref{free}: we replace the Pauli-Jordan distribution 
$
D_{m}
$
by some off-shell distribution 
$
D_{m}^{\rm off}
$
which does not verify Klein-Gordon equation but converges in some limit (in the sense of distribution theory) to
$
D_{m}.
$ 
For instance we can take 
\be
D_{m}^{\rm off} \equiv \int d\lambda \rho_{m}(\lambda) D_{\lambda}
\ee
where 
$
\rho_{m}(\lambda)
$
is some function converging, say for 
$
\lambda \rightarrow 0
$
to the distribution 
$
\delta(\lambda - m).
$
In this way all the fields from Section \ref{free} we become {\it generalized free fields} \cite{J} i.e. they will verify all properties described there except Klein-Gordon equation. The off-shell scalar field we be denoted by
$
\Phi^{\rm off}
$,
etc. However, for simplicity we will skip the index {\it off} if no confusion can arise.
 
If we keep the definion of the gauge charge unchanged we will loose the property
$
Q^{2} = 0.
$ 
If we keep unchanged the expressions of the interaction Lagrangians from the preceding Section, but replace all fields by their off-shell counterparts, we also loose the relations (\ref{descent1}). However, these relations will be replaced by 
\be
d_{Q}T^{I} = i~\partial_{\mu}T^{I\mu} + S^{I}
\label{descent1-off}
\ee
with 
$
S^{I}
$
some polynomials which will be null in the on-shell limit. We will need these expressions in the following. In the following we will denote
$
K_{c} \equiv K_{m_{c}}
$
and we assume that all fields are off-shell (we do not append the index {\it off}). We have by direct computations the following result:

\begin{thm}
The expressions
$
S^{I} 
$
have the following explicit form:
\be
S = S^{\emptyset} \equiv i~f_{abc}~u_{a}~\left(v_{b}^{\mu}~K_{c}v_{c\mu}
+ {1\over 2}~u_{a}~u_{b}~K_{c}\tilde{u}_{c}\right).
\ee
Also
\be
S^{\mu} \equiv {i\over 2}~f_{abc}~u_{a}~u_{b}~K_{c}v_{c}^{\mu}
\ee
and
\be
S^{I} = 0,\quad |I| > 1. 
\ee
\label{ym-s}
\end{thm}

Then we have:
\begin{thm}
In the off-shell formalism we can choose the the second order chronological products such that the following identity is true:
\bea
d_{Q}T^{I_{1}I_{2}}(x,y) = i {\d \over \d x^{\mu}}T^{I_{1}\mu,I_{2}}(x,y)
+ (-1)^{|I_{1}|} {\d \over \d y^{\mu}}T^{I_{1},I_{2}\mu}(x,y)
\nonumber \\
+ T(S^{I_{1}}(x),T^{I_{2}}(y)) + (-1)^{|I_{1}|}~T(T^{I_{1}}(x),S^{I_{2}}(y)).
\label{gauge-off}
\eea
\end{thm}
Indeed, if we make the substitution
$
D_{m}^{\rm off} \rightarrow D_{m}^{F,\rm off}
$
we obtain immediately the identity from (\ref{d-2}) and (\ref{descent1-off}). Similar identities are true in the higher orders of perturbation theory. Let us consider the simplest case 
$
I_{1} = I_{2} = \emptyset
$
when we have
\be
d_{Q}T(x,y) = i {\d \over \d x^{\mu}}T^{[\mu],\emptyset}(x,y)
+ {\d \over \d y^{\mu}}T^{\emptyset,[\mu]}(x,y)
+ [ T(S(x),T(y)) + (x \leftrightarrow y) ].
\label{gauge-off-empty}
\ee
Now we have a very clear origin of the anomalies. It elementary to prove that we have:
\be
T(S(x),T(y))^{\rm tree} = 
\sum_{m}~[~K_{m}D_{m}^{F,\rm off}(x - y)~A_{m}(x,y) 
+ \d_{\alpha}~K_{m}D_{m}^{F,\rm off}(x - y)~A^{\alpha}_{m}(x,y)] + \cdots
\ee
where by $\cdots$ we mean terms where the Klein-Gordon operator is acting on some off-shell field factor. So when we make the on-shell limit 
$
\lambda \rightarrow 0
$
we have 
\be
T(S(x),T(y))^{\rm tree} \rightarrow  
~\delta(x - y)~A(x,y) + \d_{\alpha}\delta(x - y)~A^{\alpha}(x,y) 
\label{A}
\ee
where the expressions
$
A(x,y)
$
and
$
A^{\alpha}(x,y)
$
are sums of the corresponding expressions
$
A_{m}(x,y)
$
and
$
A^{\alpha}_{m}(x,y)
$
respectively. In this way we have a systematic procedure to compute the tree anomalies in the second order of perturbation theory. For instance, the anomaly of the relation (\ref{gauge-off-empty}) is
\be
A(x,y) = \{\delta(x - y)~A(x,y) + [\d_{\alpha}\delta(x - y)]~A^{\alpha}(x,y)\} 
+ (x \leftrightarrow y).
\label{ano-a}
\ee

We investigate now in what conditions we can eliminate the anomaly by finite renormalizations. The first trick is to use ``partial integration" on the last terms with derivatives on the $\delta$ distribution. We obtain the equivalent form:
\be
A(x,y) = 2~\delta(x - y)~a(x,y) 
+ \left[{\d \over \d x^{\alpha}}a^{\alpha}(x,y) + (x \leftrightarrow y) \right]
\label{ano-b}
\ee
where
\be
a(x,y) \equiv A(x,y) - {\d \over \d x^{\alpha}}A^{\alpha}(x,y),
\qquad
a^{\alpha}(x,y) \equiv \delta(x - y)A^{\alpha}(x,y).
\label{a}
\ee

If we make the redefinition
\be
T(T^{\mu}(x),T(y)) \rightarrow T(T^{\mu}(x),T(y)) + i~a^{\mu}(x,y)
\ee
of the chronological products we will put the anomaly in the form 
\be
A(x,y) = 2~\delta(x - y)~a(x,x) 
\label{ano-c}
\ee

Now we have the following 
\begin{lemma}
The preceding anomaly can be eliminated iff the expression
$
a(x) = a(x,x) 
$
is a relative cocycle i.e. we have
\be
a = d_{Q}B - i \d_{\mu}B^{\mu}
\label{b}
\ee
for some Wick polynomials $B$ and
$
B^{\mu}
$.
The Wick polynomials
$
B(x)
$
and
$
B^{\mu}(x)
$
are constrained by: (a) Lorentz invariance; (b) ghost number restrictions:
\be
gh(B) = 0,\qquad gh(B^{\mu}) = 1
\ee
and (c) power counting which in our case gives:
\be
\omega(B)~,\omega(B^{\mu}) \leq 4.
\ee
\end{lemma}
The proof is very simple. Suppose that the anomaly (\ref{ano-c}) can be put in the form
\be
\delta(x - y)~a(x) =
d_{Q}R(x,y) + i {\d \over \d x^{\mu}}R^{\mu}(x,y) + {\d \over \d y^{\mu}}R^{\mu}(y,x).
\label{gauge-ano}
\ee 
with the expressions
$
R(x,y), R^{\mu}(x,y)
$
quasi-local i.e. of the form 
\be
R(x,y) = \delta(x - y)~B(x) + \cdots,\qquad
R^{\mu}(x,y) = \delta(x - y)~B^{\mu}(x)  + \cdots
\ee
where $\cdots$ are terms with higher order derivatives on the $\delta$ distribution. Then we immediately obtain from (\ref{gauge-ano}) the identity from the lemma. Conversely, if the identity from the lemma is true then we take
\be
R(x,y) = \delta(x - y)~B(x),\qquad
R^{\mu}(x,y) = \delta(x - y)~B^{\mu}(x)
\ee
and we have (\ref{gauge-ano}).

So all we have to do it to compute the expression
$
a(x,y)
$
given by the formula (\ref{a}), collapse the two variables to obtain the expression
$
a(x)
$
and impose the condition (\ref{b}). For simple models, as pure Yang-Mills theories, this computation is not very difficult but for more complicated models involving gravitation, the computation are very long and one can see the benefits of the off-shell method if one makes the comparison with the usual methods.

In the same way one can treat the other identities of the type (\ref{gauge-off}) i.e. for non-trivial sets of indices
$
I_{1},~I_{2}
$.
\newpage
\section{Second Order Gauge Invariance\label{second}}

Now we turn to the question of gauge invariance of the model in the second order of perturbation theory. The case of Yang-Mills fields has been investigated previously \cite{YM}, \cite{standard}. Using the off-shell method we have our main result:
\begin{thm}
The second order chronological products verify the gauge invariance condition 
\be
d_{Q}T(x,y) = 
i {\d \over \d x^{\mu}}T^{[\mu],\emptyset}(x,y) + i {\d \over \d y^{\mu}}T^{\emptyset,[\mu]}(x,y)
\label{t-2=int}
\ee
in the second order of perturbation theory iff the constants
$
f_{abc}
$
verify the Jacobi identity:
\be
f_{abc}f_{dec} + f_{bdc} f_{aec} + f_{dac} f_{bec} = 0, 
\label{Jacobi}
\ee

The finite renormalization of the chronological product
$
T(x,y)
$
is given by
\be
R(x,y) = \delta(x - y)~N(x)
\ee
where
\be
N = {i\over 2}~f_{abe}~f_{cde}~
v_{a}^{\mu}~v_{b}^{\nu}~v_{c\mu}~v_{d\mu}.
\ee
\end{thm}

{\bf Proof:} We will compute the anomaly using the off-shell method described in the preceding Section. The expressions 
$
A(x,y), A^{\alpha}(x,y)
$
appearing in (\ref{A}) are: 
\bea
A(x,y) = f_{abe} f_{cde} [ - u_{a}(x) v_{b\mu}(x) v_{c\nu}(y) F^{\nu\mu}_{d}(y)
+ u_{a}(x) v_{b}^{\mu}(x) u_{c}(y) \d_{\mu}\tilde{u}_{d}(y)
\nonumber \\
- 1/2~u_{a}(x) u_{b}(x) v_{c}^{\mu}(y) \d_{\mu}\tilde{u}_{d}(y)]
\eea
and
\bea
A^{\alpha}(x,y) = f_{abe} f_{cde} u_{a}(x) v_{b\nu}(x) v_{c}^{\nu}(y) v^{\alpha}_{d}(y)
\eea
respectively. Then we compute the expression (\ref{a}) and obtain:
\be
a(x,x) = d_{Q} N + 
( f_{ace}~f_{dbe} + f_{ade}~f_{bce} + f_{abe}~f_{cde} )~
(u_{a}~F_{b\mu\nu}~v_{c}^{\mu}~v_{d}^{\nu} - u_{a}~u_{b}~v_{c}^{\mu}~\partial_{\mu}\tilde{u}_{c} ) 
\ee
where $N$ is the expression from the statement. If we impose the condition (\ref{b}) taking an arbitrary ansatz for $B$ and
$
B^{\mu}
$  
we obtain that the last term in the right hand side must be null i.e. we have Jacobi identity. 
$\qed$

\begin{rem} If we substitute the renormalized expression of the chronological product
$$
T^{R}(x,y) \equiv T(x,y)+ \delta(x - y)~N(x)
$$
in the $S$-matrix (\ref{S-matrix}) then we formally obtain the full (classical) Yang-Mills Lagrangian: the tri-linear part is given by the first order chronological product (\ref{int-ym}) and the quadratic part by the finite renormalization $N$. 
\end{rem}

We can extend the argument for the general second order chronological products: they verify the gauge invariance condition (\ref{t-2}) in the second order of perturbation theory {\it iff} the constants verify the Jacobi identity and we have finite renormalizations of the chronological product
$
T^{I_{1},I_{2}}(x,y)
$
are given by the following expressions
\be
R^{I_{1},I_{2}}(x,y) = \delta(x - y)~N^{I_{1},I_{2}}(x)
\ee
where
\bea
N^{\emptyset\emptyset} = {i\over 2}~f_{abe}~f_{cde}~
v_{a}^{\mu}~v_{b}^{\nu}~v_{c\mu}~v_{d\mu}
\nonumber\\
N^{[\mu]\emptyset} = - i~f_{abe}~f_{cde}~
u_{a}~v_{b}^{\nu}~v_{c}^{\nu}~v_{d}^{\mu}
\nonumber\\
N^{[\mu][\nu]} = - i~f_{abe}~f_{cde}~
( u_{a}~v_{b}^{\nu}~u_{c}~v_{d}^{\mu} - \eta^{\mu\nu}~u_{a}~v_{b}^{\nu}~u_{c}~v_{d\nu} )
\nonumber\\
N^{[\mu\nu]\emptyset} = - {i\over 2}~f_{abe}~f_{cde}~
u_{a}~u_{b}~v_{c}^{\mu}~v_{d}^{\nu}
\nonumber\\
N^{[\mu\nu][\rho]} = - {i\over 2}~f_{abe}~f_{cde}~
[ \eta^{\mu\rho}~u_{a}~u_{b}~u_{c}~v_{d}^{\nu} - (\mu \leftrightarrow \nu ) ]
\nonumber\\
N^{[\mu\nu][\rho\sigma]} = {i\over 4}~f_{abe}~f_{cde}~
(\eta^{\mu\rho}~\eta^{\nu\sigma} - \eta^{\nu\rho}~\eta^{\mu\sigma})~u_{a}~u_{b}~u_{c}~u_{d}.
\eea

\section{Conclusions}

The preceding result can be extended to the most general case of Yang-Mills fields (massless and massive), Dirac fields and the gravitational field in interaction \cite{general}. The elimination of the anomalies in higher orders of perturbation theory is a very complicated problem and, for the moment, it can be done only for special cases like QED and related models, where we have a special new symmetry (charge conjugation) \cite{ano}, \cite{caciulata3}. 

\vskip 1cm
{\bf Acknowledgment:}  This work was supported by CNCSIS-UEFISCSU, project number PNII - IDEI 454/2009, code CNCSIS ID-44 and by ANCS project number PN 09 37 01 02/2009.

\end{document}